# A valleytronic diamond transistor: electrostatic control of valley-currents and charge state manipulation of NV centers

N. Suntornwipat[1], S. Majdi[1], M. Gabrysch[1], K. K. Kovi[1,3], V. Djurberg[1], I. Friel[2], D. J. Twitchen[2] and J. Isberg[1*]

[1] *Division for Electricity, Department of Electrical Engineering, Uppsala University, Box 65, 751 03, Uppsala, Sweden.*
[2] *Element Six, Global Innovation Centre, Fermi Ave, Harwell Oxford, Oxfordshire OX11 0QR, United Kingdom.*
[3] *Center for Nanoscale Materials, Argonne National Laboratory, Argonne, IL-60439, United States.*

(Dated: 17nd Sep 2020)

The valley degree of freedom in many-valley semiconductors provides a new paradigm for storing and processing information in valleytronic and quantum-computing applications. Achieving practical devices require all-electric control of long-lived valley-polarized states, without the use of strong external magnetic fields. Attributable to the extreme strength of the carbon-carbon bond, diamond possesses exceptionally stable valley states which provides a useful platform for valleytronic devices. Using ultra-pure single-crystalline diamond, we here demonstrate electrostatic control of valley-currents in a dual gate field-effect transistor, where the electrons are generated with a short UV pulse. The charge- and the valley- current measured at receiving electrodes are controlled separately by varying the gate voltages. A proposed model based on drift-diffusion equations coupled through rate terms, with the rates computed by microscopic Monte Carlo simulations, is used to interpret experimental data. As an application, valley-current charge-state modulation of nitrogen-vacancy (NV) centers is demonstrated.

Charge and spin are both well-defined quantum numbers in solids. The same is true for the valley pseudospin, also known as valley polarization, in various multi-valley semiconductor materials. The valley pseudospin degree of freedom presents an attractive resource for information processing, a subject that has been termed *valleytronics* in analogy with spintronics for spin-based technology. The topic of valleytronics has attracted considerable attention lately because of its fundamental interest to the physics community. Due to the large separation in momentum space between valleys in certain materials, the valley pseudospin can be robust against lattice deformation and low-energy phonon scattering. Creation and detection of valley-polarized electrons have been achieved in several materials, such as in AlAs where valley polarization was induced by a symmetry-breaking strain,[1] in $MoS_2$ by means of circularly polarized light[2–4] and in bulk bismuth by using a rotating magnetic field to modulate the different contributions of different valleys.[5] Previously, it has been demonstrated that valley-polarized states can be created in diamond.[6] This occurs by the hot electron repopulation effect in a high electric field or by separating an initially unpolarized population into differently polarized electron states by means of crossed electric and magnetic fields, i.e., by the Hall effect.

However, to realize practical valleytronic applications, establishing a fast, scalable and direct electrical control of valley states is crucial.[7] So far, electrical control of valley transport has not been studied experimentally outside the field of topological transport in low-dimensional materials. In this article, we demonstrate such electrical control of valley transport in double-gated diamond field-effect transistors (FET), where a short UV pulse is used to generate the valley-polarized electrons. We find that it is possible to control the charge- and valley- current separately at receiving electrodes, thereby enabling rapid and scalable valley-current control. We also demonstrate charge-state modulation of nitrogen-vacancy (NV) centers in diamond through valley-currents. NV centers in diamond have important applications in e.g. single-spin magnetometry[8] and single photon sources[9] Our transistors show that it is possible to manipulate the charge states of NV centers locally in devices solely by application of a gate bias voltage.

*Corresponding author: Prof. Jan Isberg
Email: jan.isberg@angstrom.uu.se

Diamond is unique among solids as the extreme rigidity of the lattice leads to suppressed intervalley scattering and highly stable valley states.[10] Diamond also exhibits many other valuable properties for solid-state QIP, e.g. a high electron mobility[11] and the existence of the NV center with ultra-long spin coherence time.[12] The conduction band structure of diamond is similar to that of silicon, with six equivalent conduction band valleys oriented along the {100} axes. The minima are symmetrically situated at 76% of the distance from the □ point to the X point as indicated in the energy diagram in Figure 1(a). At temperatures below ~ 80 K, the intervalley scattering rates become so low ($< 10^4 \text{ s}^{-1}$) that electrons are effectively confined to a given valley.[10] Therefore, electrons exist in six different valley pseudospin states. Valley electrons have a large ratio between longitudinal effective mass ($m_\parallel$) and transversal effective mass ($m_\perp$), i.e. $m_\perp \approx 5.5\ m_\parallel$.[13] This, in conjunction with the freeze-out of electrons into different valleys at low temperatures, leads to a break of the cubic symmetry of the crystal and to anisotropic charge transport properties that depend on the specific value of the pseudospin. This dependence makes it possible to separate electrons with different pseudospins by their drift in a suitably arranged electric field, as illustrated in Figure 1(a). The anisotropic energy dispersion results in a conduction anisotropy, different for electrons in valleys on different axes. In an applied field ($\bar{E}$), electrons in separate valleys acquire different drift velocities ($\bar{v}$). If intervalley scattering rates are small, this results in a separation of the electron populations.

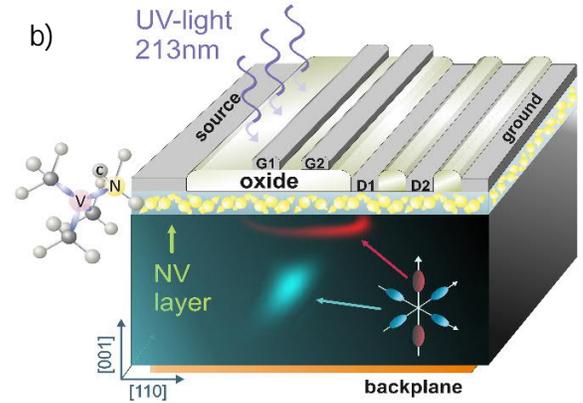

**Figure 1.** (a) The conduction band structure of diamond. The central image of the left figure depicts the first Brillouin zone, together with iso-energy surfaces surrounding the six conduction band minima ("valleys"). The blue dot in the band structure plot (upper left) indicates the position of one conduction band minimum. The remaining images illustrate the energy dispersion in different valleys. As different valleys have different anisotropic effective mass tensors, electrons in different valleys drift in different directions when an electric field is applied. (b) A schematic of a dual-gate two-drain transistor together with a simulation showing the electron density in different valleys a certain time after illumination. G1 and G2 are two independently controllable gates and D1 and D2 are the two drains. A thin ion-implanted layer with NV centers is incorporated. Electroluminescence from the NV centers can be observed through a semitransparent back contact.

Diamond FET devices on selected single-crystalline diamond plates were made to demonstrate electrostatic control of valley pseudospin states. These plates were synthesized by Element Six Ltd. with ultra-low nitrogen impurity concentration, below 0.05 ppb. FET:s in diamond have been profusely studied in recent years[14] but these have however not been previously employed to study valley polarized currents. The designed transistor is depicted in Figure 1(b) and comprises a source electrode, two gate electrodes isolated from the diamond by a 30 nm $Al_2O_3$ dielectric layer and two drain electrodes at which induced currents can be monitored. The $Al_2O_3$ layer serves both as an insulating gate oxide and as a surface passivation layer reducing the surface scattering rate and the interface trap density. The reason for the multiple drain electrodes is that we wish to show that valley-polarized currents can be directed to different locations in the crystal. The double-gate configuration allows for more control of the valley-currents than a single-gate configuration does. Holding one gate at a constant voltage while varying the bias on the second gate offer improved ability to separate the charge clouds with precision. With the chosen geometry, drain currents are only induced when charges are drifting in close proximity to the contacts. We emphasize, in contrast to (Ref. 15-18), that the electrostatic potential at the gate acts directly through coulomb interaction and is not used to modulate the energy gap or affect the Berry phase. Details of the device processing are described in the supporting information.

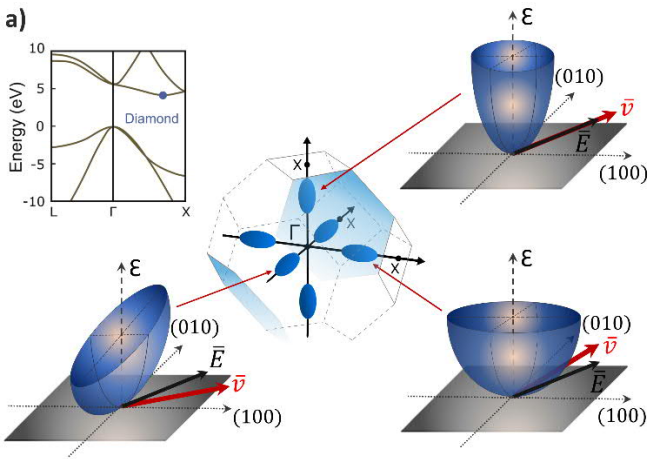

The charge transport was modelled using drift-diffusion and Poisson equations with several electron concentrations, one for each valley in the Brillouin zone (BZ). As valleys on the same axis in the BZ have the same effective mass tensor (to quadratic order) the contribution from these valley-pairs can be added for simplicity, resulting in three different electron concentrations $n^l$ ($l = 1,2,3$). For low electric fields, the average carrier drift velocity for electrons in the $l$:th valley, $\bar{v}^l$, is proportional to the electric field $\bar{E}$, i.e. $\bar{v}^l = \boldsymbol{\mu}^l \bar{E}$ for the $l$:th valley-pair. Here, the mobility tensor $\boldsymbol{\mu}^l$ is related to the effective mass tensor $\boldsymbol{M}^l$ by $\boldsymbol{\mu}^l \equiv q <\tau> (\boldsymbol{M}^l)^{-1}$, where $q$ is the elementary charge and $<\tau>$ is the average intravalley relaxation time. Since $\boldsymbol{\mu}^l$ is a tensor, the velocity $\bar{v}^l$ and the electric field $\bar{E}$ are not parallel in general. As $\bar{v}^l$ depends on the valley index, electrons with different valley pseudospin tend to drift in different directions. The effective mass tensor $\boldsymbol{M}^l$ is given by the curvature of the conduction band $\mathcal{E}(\bar{k})$ at the respective band minima, $(\boldsymbol{M}^1)_{\alpha\beta} = \frac{1}{\hbar^2}\frac{\partial^2 \mathcal{E}(\bar{k})}{\partial k_\alpha \partial k_\beta}$. For a cubic semiconductor with the energy minima on the principal axes in k-space, such as diamond, this gives:

$$\boldsymbol{\mu}^1 = \begin{pmatrix} \mu_\| & 0 & 0 \\ 0 & \mu_\perp & 0 \\ 0 & 0 & \mu_\perp \end{pmatrix}, \boldsymbol{\mu}^2 = \begin{pmatrix} \mu_\perp & 0 & 0 \\ 0 & \mu_\| & 0 \\ 0 & 0 & \mu_\perp \end{pmatrix},$$

$$\boldsymbol{\mu}^3 = \begin{pmatrix} \mu_\perp & 0 & 0 \\ 0 & \mu_\perp & 0 \\ 0 & 0 & \mu_\| \end{pmatrix} \quad (1)$$

Where the ratio between transversal mobility $\mu_\perp$ and longitudinal mobility $\mu_\|$ equals the inverse ratio of the corresponding effective masses, i.e., $\mu_\perp \approx 5.5 \mu_\|$,[13] resulting in a strong charge transport anisotropy. Intervalley *f*-scattering is incorporated in the model by including rate terms in the drift-diffusion equations. In contrast, *g*-scattering is inconsequential as it occurs between valleys on the same axis. Under these assumptions and with the convention that summation is implied for repeated Greek indices, the drift-diffusion and Poisson equations read:

$$\frac{\partial n^l}{\partial t} = -\boldsymbol{\mu}^l_{\alpha\beta}\nabla_\alpha(n^l \nabla_\beta \phi) + \frac{k_B T_c(\nabla\phi)\boldsymbol{\mu}^l_{\alpha\beta}}{q}\nabla_\alpha\nabla_\beta n^l$$

$$+ \frac{\Gamma(\nabla\phi)}{2}\left(\sum_{m=1}^{3} n^m - 3n^l\right), l = 1,2,3$$

$$\nabla^2 \phi = \frac{q}{\varepsilon}\sum_{l=1}^{3} n^l \quad (2)$$

Here, $\phi$ is the electrostatic potential, $\Gamma$ is the E-field dependent intervalley relaxation time (Ref. 19), $k_B$ is Boltzmann's constant and $T_c$ is the carrier temperature which is also E-field dependent. The dependence of $\Gamma$ and $T_c$ on the electric field was in the finite element method (FEM) simulation treated by creating a look-up table obtained from Monte Carlo (MC) simulations in conjunction with an interpolation scheme (the MC model is described in the supporting information).

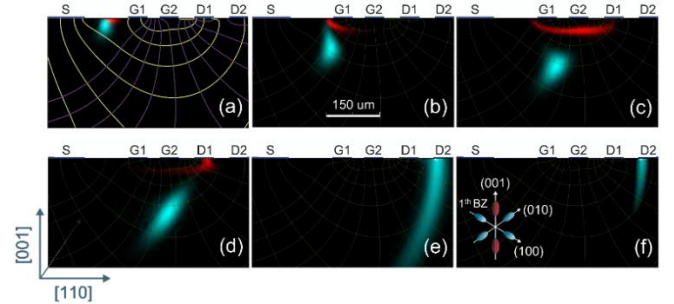

**Figure 2.** Simulation of valley pseudospin transport in a double-gated transistor. The simulation (a-f) shows the electron density in different valleys ((001) valleys in red color and (010) and (100) valleys in cyan) and at different times (2.5 ns, 8 ns, 13.5 ns, 19 ns, 41.5 ns and 52 ns respectively) after current injection. The parameters of the applied bias are: $V_{source} = -7$ V, $V_{G1} = -5$ V, $V_{G2} = -2.4$ V, $V_{backplane} = -2.4$ V, with the drains grounded. The inhomogeneous electric field distribution is in (a) indicated by purple equipotential lines and yellow field lines.

An example of a drift-diffusion simulation of the transistor depicted in Figure 1(b) is shown in Figure 2. Charge is injected at the edge of the source contacts, and states of different valley pseudospin can be seen to drift to different drain contacts. The simulation show for a specific set of bias voltages the entire process from generation of the charge, drift and finally detection of the electrons. For this choice of bias voltages, electrons in (001) valleys predominantly drift to D1 while electrons in other valleys are collected at D2. To further investigate this, we recorded the simulated current at drains D1 and D2 for a situation where the voltage at G1 was varied and all other bias voltages were held fixed. It can be noted that the current reaches two separate peaks at different times and at different gate voltages. The contribution to the current from different valley pseudospins is indicated in Figure 3(a,b) by red and blue contours. It is clear that the two peaks originate from electrons in different pseudospin states. In this case, the electrons in the (001) valleys travel faster and arrive at the drain contacts earlier than the electrons in the (010) and (100) valleys.

With a suitable choice of gate voltage $V_{G1}$ it is possible to direct electrons in the (010) and (100) valleys to one contact and (001) electrons to another. Figure 3(c) shows a simulation of how the degree of valley pseudospin arriving at the second drain (D2) can be modulated by varying the gate voltage $V_{G1}$. This demonstrates theoretically that in our devices the (001) valleys' contribution to the total charge can be modulated with a very high fidelity, from 10% to 90%.

In our first experiment, the edge of the transistor source contact was illuminated using a passively Q-switched 213 nm wavelength laser with a repetition rate of 300 Hz. This was done to measure the charge drift time and to compare it with the simulation. To reduce electron-electron scattering to negligible levels, the pulse energy was limited to < 1 nJ/pulse by an attenuator resulting in a peak carrier concentration $< 10^{10}$ cm$^{-3}$. Below this concentration the electric field is influenced by less than 5% by the space charge in the electron clouds. A low power used, < 0.3 µW, ensures negligible sample heating. Electron-hole pairs were created near the edge of the source electrode by short pulses (800 ps) of photons with above-bandgap energy ($hv = 5.82$ eV > $E_{gap} = 5.47$ eV). The optical excitation creates electron-hole pairs, with the electrons equally populating the six valleys. By applying different negative bias voltages at the source, the gate and the back contacts, the electrons drift towards the drain contacts with different velocities and directions depending on their valley polarization. The holes are rapidly extracted at the source electrode and their contribution to the drain current is negligible. The induced currents were measured at the drain contacts, which were held at (virtual) ground potential. Further experimental details are given in the supporting information.

Figure 4 shows measured time-resolved currents from the transistor where we have applied the same set of bias voltages as in the simulation described earlier. The currents originating from longitudinal and transversal valleys are observed as two distinguished peaks and are highlighted by the dashed lines. These two current peaks can be observed for different gate voltages and also at different times. The behavior of the measured currents is well reproduced in the carrier transport simulations, as can be noted by comparing with Figure 3. The difference in carrier arrival times shows that these peaks can be attributed to electrons in different states of valley pseudospin arriving at the drain contacts.

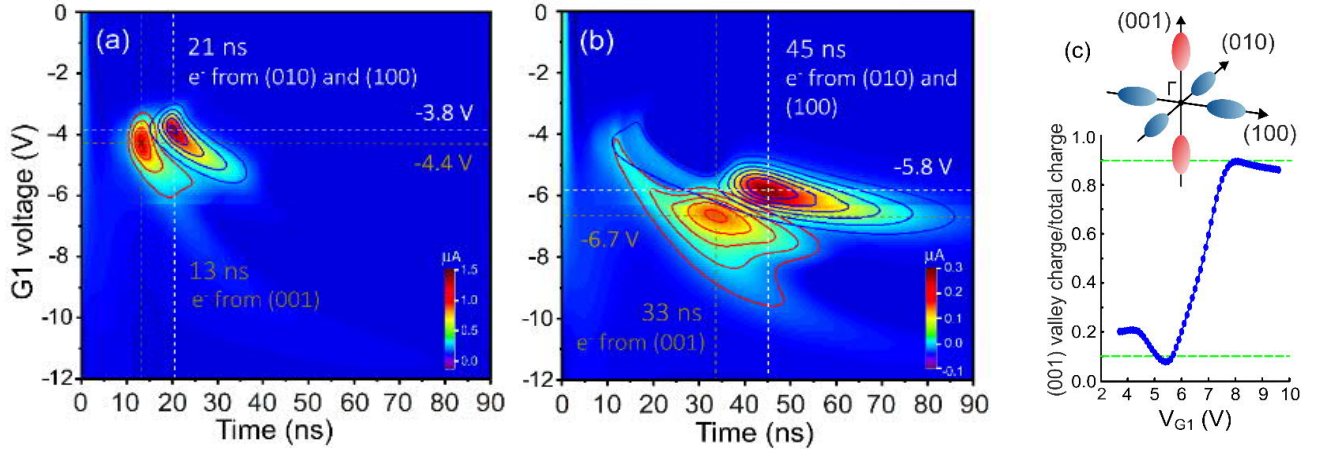

**Figure 3.** Color plot of the simulated time-resolved total induced currents in (a) D1 and (b) D2. The voltage at G1 is varied between 0 and −12 V in steps of 0.2 V, and the other bias voltages are fixed: $V_{source} = -7$ V, $V_{G2} = -2.4$ V, $V_{backplane} = -2.4$ V and drains grounded. The superimposed red and blue contours specify the contribution to the total current from different valleys. (c) The simulated (001) valley contribution to the total charge collected at drain D2, for the same bias voltages as for (a) and (b), at 78 K.

Experimentally observed arrival times are within 8% of the values obtained from the simulation. The absolute values of the gate voltages at peak current are slightly underestimated in the simulation (5-20%) which can be explained by the presence of interface states that are not accounted for in the simulation. The main difference between the experimental and the simulated data is that the peaks are broader experimentally. This is presumably due to small inhomogeneities in the electric field originating from extended defects such as dislocation bundles.

In the second experiment we have used our transistors to achieve charge-state manipulation of NV centers in diamond. This was done by adding a thin (120 nm) layer with NV centers below the top electrodes as in Figure 1(b). This layer was added to demonstrate the charge-state manipulation and to make it possible to optically monitor current densities by electroluminescence (EL).[20] The EL was observed through the sample and the semitransparent back contact using a home-built microscope equipped with a cooled sCMOS camera. The rear surface was covered with a thin gold layer providing an optically semitransparent back contact, which provides a well-defined reference potential, see more details in supporting information.

In this experiment electron-hole pairs were generated near the source electrode by the 213 nm wavelength laser at a repetition rate of 660 Hz and a pulse energy of 10 nJ/pulse. The sample was from below imaged with a CCD camera through the semitransparent contact. Exposure times of several seconds ensured that enough light was collected to yield clear images. Figure 5(b,c) shows how the observed electroluminescence shifts between different locations in the device solely by varying the gate voltages. These images are composites with two different exposure times, 60 ms inside the dashed yellow circle and 8 s outside. The line inside the circle is strong luminescence from the region where electron-hole pairs were generated. The EL was observed at D1 for $V_{G1} = -9.6$ V and $V_{G2} = -5.2$ V (Figure 5b), while it clearly shifted to D2 by changing the gate bias to $V_{G1} = -13$ V and $V_{G2} = -7.0$ V (Figure 5c). In Figure 5(b,c), the EL spectrum comes from the rightmost part of the device (to the right of G2). The illuminated area was masked off so that light from this region did not enter the spectrometer and the measurements were integrated over time. The conduction band electrons that are generated by illumination (213 nm laser pulses) are excited across the indirect band gap with a possible (many orders of magnitude smaller) contribution from photoionization of NV centers. The presence of a dominant 575 nm peak in the luminescence spectra together with several phonon replicas (Figure 5d) show that the luminescence indeed originates from neutral charge ($NV^0$) centers. The specific optical signature of the different charge states of NV centers has been discussed and identified in (Ref. 9,21-23). The EL can be understood[24] by a three stage process: (i) The $NV^0$ center traps a conduction band electron and is converted to the negative charge state ($NV^-$). (ii) The $NV^-$ center reverts to the neutral state by de-trapping the electron, either by direct tunneling to the contact or by hopping conduction via other defects, leaving the $NV^0$ center in an exited state. (iii) The $NV^0$ center reverts to the ground state and emits a photon.

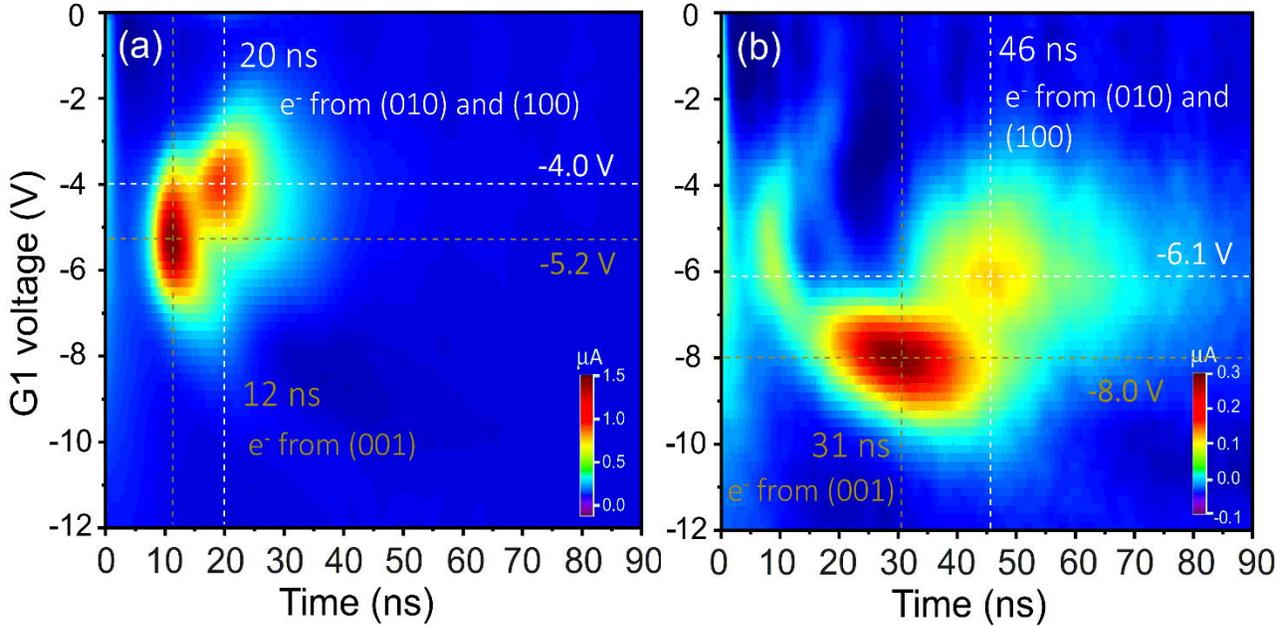

**Figure 4.** Experimentally observed time-resolved currents in a double-gated transistor. Time-resolved drain currents, (a) at D1 and (b) at D2, measured in a double-gated transistor at 78 K as a function of applied gate voltage, $V_{G1}$, and time after carrier injection. The same set of bias voltages used in the simulations has been applied in the experiment: $V_{\text{source}} = -7$ V, $V_{G2} = -2.4$ V, $V_{\text{backplane}} = -2.4$ V and drains grounded.

In summary, we have demonstrated electrostatic control of valley-currents in double-gated diamond field-effect transistors as well as charge-state manipulation of NV centers. This was done by detecting different electron drift times in time-resolved current measurements and also by direct observation of electroluminescence from NV centers near the receiving contacts. Transport simulations of valley-polarized electrons show good agreement with the observations. These valleytronic devices enable electrostatic manipulation of valley-currents and can be used to deliver electrons with a high degree of valley polarization for, e.g., electrical pumping of color centers in diamond for single photon sources. They offer a possible solid state platform for valleytronic-based information processing and for further investigations into the physics of spin-valley states. We anticipate that such devices will play a significant role in quantum information processing and future quantum computing.

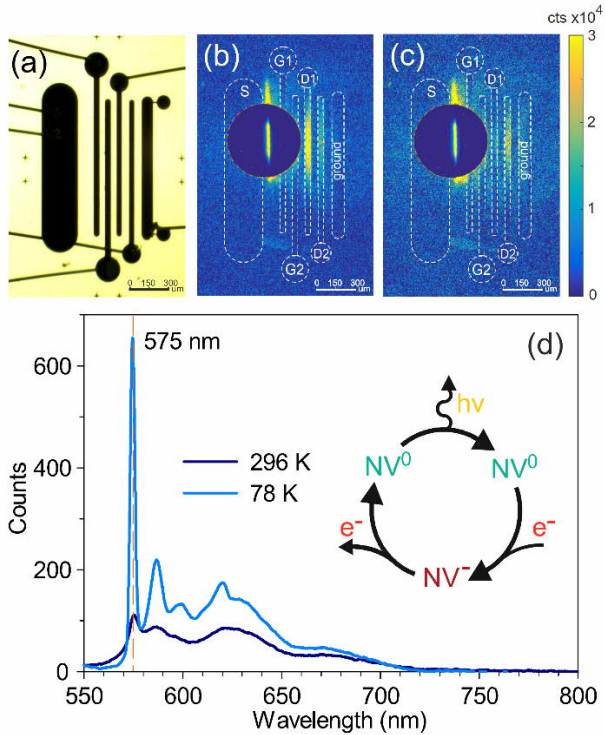

**Figure 5.** Electroluminescence of the valley pseudospin current. (a) Transmitted light optical micrograph of the transistor. (b,c) Luminescence images at 78 K for different gate voltages, with the transistor imaged from above. The transistor is near the contact surface equipped with a thin layer of NV centers that luminesce with a typical $NV^0$ spectral signature, with a strong zero phonon line (ZPL) at 575 nm at 78 K, together with associated phonon replicas, as shown in (d). For comparison, the room-temperature spectrum is also presented in (d). The spectrum is made with the leftmost part (left of G2) masked off. The position of the contacts are indicated in (b,c) by dashed white lines. Note the shift in electroluminescence from D1 to D2 as the gate bias is varied.

**ACKNOWLEDGMENT**

This study is supported by the Swedish Research Council (research grant 2018-04154), the ÅForsk Foundation (Grant No. 15-288 and 19-427), the Olle Engkvists Foundation (198-0384) and the STandUP for Energy strategic research framework. The Monte Carlo simulations were performed on resources provided by the Swedish National Infrastructure for Computing (SNIC) through the Uppsala Multidisciplinary Center for Advanced Computational Science.

# REFERENCES

(1) Gunawan, O.; Shkolnikov, Y. P.; Vakili, K.; Gokmen, T.; De Poortere, E. P.; Shayegan, M. Valley Susceptibility of an Interacting Two-Dimensional Electron System. *Phys. Rev. Lett.* **2006**, *97* (18), 186404.

(2) Cao, T.; Wang, G.; Han, W.; Ye, H.; Zhu, C.; Shi, J.; Niu, Q.; Tan, P.; Wang, E.; Liu, B.; Feng, J. Valley-Selective Circular Dichroism of Monolayer Molybdenum Disulphide. *Nat Commun* **2012**, *3*, 887.

(3) Mak, K. F.; He, K.; Shan, J.; Heinz, T. F. Control of Valley Polarization in Monolayer $MoS_2$ by Optical Helicity. *Nat Nano* **2012**, *7* (8), 494.

(4) Zeng, H.; Dai, J.; Yao, W.; Xiao, D.; Cui, X. Valley Polarization in $MoS_2$ Monolayers by Optical Pumping. *Nat Nano* **2012**, *7* (8), 490.

(5) Zhu, Z.; Collaudin, A.; Fauque, B.; Kang, W.; Behnia, K. Field-Induced Polarization of Dirac Valleys in Bismuth. *Nat Phys* **2012**, *8* (1), 89.

(6) Isberg, J.; Gabrysch, M.; Hammersberg, J.; Majdi, S.; Kovi, K. K.; Twitchen, D. J. Generation, Transport and Detection of Valley-Polarized Electrons in Diamond. *Nature materials* **2013**, *12* (8), 760.

(7) Ang, Y. S.; Yang, S. A.; Zhang, C.; Ma, Z.; Ang, L. K. Valleytronics in Merging Dirac Cones: All-Electric-Controlled Valley Filter, Valve, and Universal Reversible Logic Gate. *Phys. Rev. B* **2017**, *96* (24), 245410.

(8) Balasubramanian, G.; Charrf, Y.; Kolesov, R.; Al-Hmoud, M.; Tisler, J.; Shin, C.; Kim, C.; Wojcik, A.; Hemmer, P. R.; Krueger, A.; Hanke, T.; Leitenstorfer, A.; Bratschitsch, R.; Jelezko, F.; Wrachtrup, J. Nanoscale Imaging Magnetometry with Diamond Spins under Ambient Conditions. *Nat. (UK)* **2008**, *455* (7213), 648.

(9) Schreyvogel, C.; Polyakov, V.; Wunderlich, R.; Meijer, J.; Nebel, C. E. Active Charge State Control of Single NV Centres in Diamond by In-Plane Al-Schottky Junctions. *Scientific Reports* **2015**, *5* (1), 12160.

(10) Hammersberg, J.; Majdi, S.; Kovi, K. K.; Suntornwipat, N.; Gabrysch, M.; Isberg, J. Stability of Polarized States for Diamond Valleytronics. *Applied Physics Letters* **2014**, *104*, 232105.

(11) Isberg, J.; Hammersberg, J.; Johansson, E.; Wikström, T.; Twitchen, D. J.; Whitehead, A. J.; Coe, S. E.; Scarsbrook, G. A. High Carrier Mobility in Single-Crystal Plasma-Deposited Diamond. *Science* **2002**, *297* (5587), 1670.

(12) Balasubramanian, G.; Neumann, P.; Twitchen, D.; Markham, M.; Kolesov, R.; Mizuochi, N.; Isoya, J.; Achard, J.; Beck, J.; Tissler, J. Ultralong Spin Coherence Time in Isotopically Engineered Diamond. *Nature materials* **2009**, *8* (5), 383.

(13) Naka, N.; Fukai, K.; Handa, Y.; Akimoto, I. Direct Measurement via Cyclotron Resonance of the Carrier Effective Masses in Pristine Diamond. *Physical Review B* **2013**, *88* (3), 035205.

(14) Donato, N.; Rouger, N.; Pernot, J.; Longobardi, G.; Udrea, F. Diamond Power Devices: State of the Art, Modelling, Figures of Merit and Future Perspective. *J. Phys. D: Appl. Phys.* **2019**, *53* (9), 093001.

(15) Ju, L.; Shi, Z.; Nair, N.; Lv, Y.; Jin, C.; Velasco Jr, J.; Ojeda-Aristizabal, C.; Bechtel, H. A.; Martin, M. C.; Zettl, A.; Analytis, J.; Wang, F. Topological Valley Transport at Bilayer Graphene Domain Walls. *Nature* **2015**, *520* (7549), 650.

(16) Sui, M.; Chen, G.; Ma, L.; Shan, W.-Y.; Tian, D.; Watanabe, K.; Taniguchi, T.; Jin, X.; Yao, W.; Xiao, D.; Zhang, Y. Gate-Tunable Topological Valley Transport in Bilayer Graphene. *Nat Phys* **2015**, *11* (12), 1027.

(17) Shimazaki, Y.; Yamamoto, M.; Borzenets, I. V.; Watanabe, K.; Taniguchi, T.; Tarucha, S. Generation and Detection of Pure Valley Current by Electrically Induced Berry Curvature in Bilayer Graphene. *Nat Phys* **2015**, *11* (12), 1032.

(18) Li, J.; Wang, K.; McFaul, K. J.; Zern, Z.; Ren, Y.; Watanabe, K.; Taniguchi, T.; Qiao, Z.; Zhu, J. Gate-Controlled Topological Conducting Channels in Bilayer Graphene. *Nat Nano* **2016**, *11*, 1060.

(19) Hammersberg, J.; Majdi, S.; Kovi, K. K.; Suntornwipat, N.; Gabrysch, M.; Twitchen, D. J.; Isberg, J. Stability of Polarized States for Diamond Valleytronics. *Applied Physics Letters* **2014**, *104* (23), 232105.

(20) Kato, H.; Wolfer, M.; Schreyvogel, C.; Kunzer, M.; Müller-Sebert, W.; Obloh, H.; Yamasaki, S.; Nebel, C. Tunable Light Emission from Nitrogen-Vacancy Centers in Single Crystal Diamond PIN Diodes. *Appl. Phys. Lett.* **2013**, *102* (15), 151101.

(21) Mizuochi, N.; Makino, T.; Kato, H.; Takeuchi, D.; Ogura, M.; Okushi, H.; Nothaft, M.; Neumann, P.; Gali, A.; Jelezko, F.; Wrachtrup, J.; Yamasaki, S. Electrically Driven Single-Photon Source at Room Temperature in Diamond. *Nature Photonics* **2012**, *6* (5), 299.

(22) Bourgeois, E.; Gulka, M.; Nesladek, M. Photoelectric Detection and Quantum Readout of Nitrogen-Vacancy Center Spin States in Diamond. *Advanced Optical Materials* **2020**, *8* (12), 1902132.

(23) Siyushev, P.; Nesladek, M.; Bourgeois, E.; Gulka, M.; Hruby, J.; Yamamoto, T.; Trupke, M.; Teraji, T.; Isoya, J.; Jelezko, F. Photoelectrical Imaging and Coherent Spin-State Readout of Single Nitrogen-Vacancy Centers in Diamond. *Science* **2019**, *363* (6428), 728.

(24) Fedyanin, D. Y.; Agio, M. Ultrabright Single-Photon Source on Diamond with Electrical Pumping at Room and High Temperatures. *New J. Phys.* **2016**, *18* (7), 073012.

# Supporting Information

*Monte Carlo simulations:*
To compute the electric field dependence of the intervalley relaxation time and the carrier temperature Monte Carlo simulations were performed. Because of the low concentration of impurities in our samples impurity scattering was neglected and only electron-phonon interactions that are allowed by symmetry selection rules were included in the simulations. Likewise, carrier-carrier scattering was neglected due to the low carrier concentrations considered. We use a simple conduction band structure, consisting of six parabolic but anisotropic (ellipsoidal) valleys. This simplification is adequate at the moderate electric fields considered in this work and it reduces the computational effort considerably, compared to full-band simulations. Scattering by acoustic phonons is treated through inelastic deformation-potential interaction. In a crystal with the valleys centered on the {100} axes, the deformation potential tensor has two independent components for interactions with transversal and longitudinal phonon modes. However, the effect of this anisotropy is small so as a further simplification it is possible to average over the phonon wave vector angle, which leaves only one independent component of the deformation potential tensor. With these simplifications the scattering rate $P_{ac}$ per unit volume can be written as:[1]

$$P_{ac}(\bar{k}, \bar{k} \pm \bar{\kappa}) = \frac{q\pi D_A^2}{\rho v} \left\{ \left( \exp(\frac{\hbar|\bar{\kappa}|v}{k_B T}) - 1 \right)^{-1} + \tfrac{1}{2} \mp \tfrac{1}{2} \right\} \cdot \delta(E_c(\bar{k} \pm \bar{\kappa}) - E_c(\bar{k}) \mp \hbar|\bar{\kappa}|v) \qquad (S1)$$

where $\bar{\kappa}$ is the phonon wave vector and $\bar{k}$ and $\bar{k} \pm \bar{\kappa}$ are the initial and final state electron wave vectors, respectively. The upper sign is used for phonon absorption and the lower sign for phonon emission. $v$ is an averaged sound velocity $v = (2v_t + v_l)/3$ and $\rho$ is the material density (3.515 g/cm$^3$). The acoustic deformation potential $D_A$ is assumed to be 12.0 eV[2]. Intervalley scattering is included in the model through *f*- and *g*-scattering deformation potential interactions, $D_f = D_g = 4\times10^8$ eV/cm and we assume a constant transition phonon energy $\hbar\omega_f$ = 110 meV, $\hbar\omega_g$ = 165 meV. The *f*- and *g*-scattering rates per unit volume are given by:[1]

$$P_{f,g}(\bar{k}, \bar{k} \pm \bar{\kappa}) = \frac{4\pi D_{f,g}^2}{\rho \omega_{f,g}} \left\{ \left( \exp(\frac{\hbar\omega_{f,g}}{k_B T}) - 1 \right)^{-1} + \tfrac{1}{2} \pm \tfrac{1}{2} \right\} \cdot \delta(E_c(\bar{k} \pm \bar{\kappa}) - E_c(\bar{k}) \pm \hbar\omega_{f,g}) \qquad (S2)$$

For the final state selection in the Monte Carlo simulation we keep to the treatment described in (Ref. 1). The simulated dependence of the *f*-scattering rate and the electron temperature on the electric field is plotted in Figure S1. These are used together with an interpolation scheme in the drift-diffusion charge transport simulations.

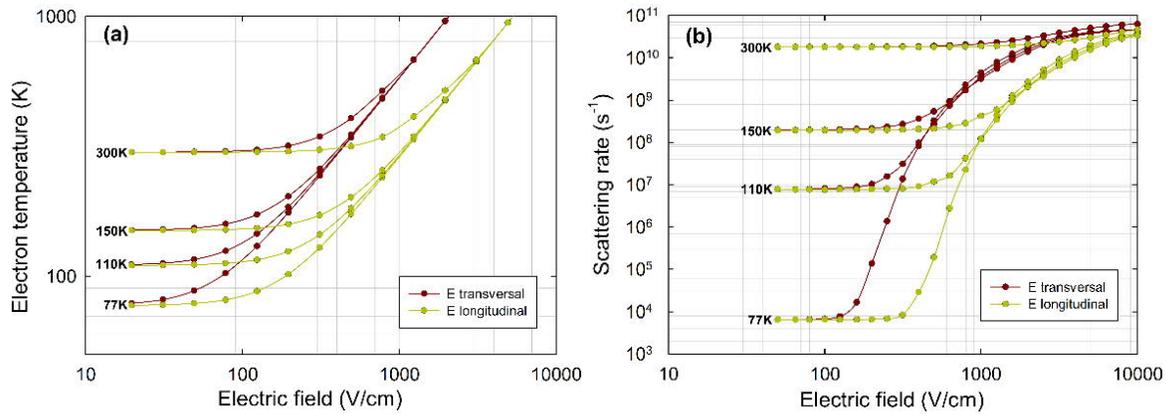

**Figure S1.** Electron temperature and *f*-scattering rate vs. E-field from Monte Carlo simulations. (a) Data shown for four different lattice temperatures and for the E-field in parallel with longitudinal (yellow circles) and transversal (red circles) to the major axis of the valley energy ellipsoid. As the E-field increases, the lattice temperature increases more rapidly for the electrons on the transversal axis. (b) Illustrates the scattering rate of the electrons in different valleys parallel or orthogonal to the E-field. The simulation results reveal a higher scattering rate ratio between the transversal and longitudinal valleys at lower temperatures.

## *Sample processing:*

The transistors were made using freestanding single-crystalline CVD (001) plates synthesized under conditions of high purity by Element Six Ltd. The samples are $4.5 \times 4.5$ mm, with thicknesses ranging from 390 to 510 µm. In our experiments, the influence of ionized impurity scattering is minimized by selecting the purest CVD diamond samples available. The concentration of the dominant impurity, nitrogen, is below $10^{13}$ cm$^{-3}$ in all samples, as determined by electron paramagnetic resonance (EPR). From previous studies[3] it is known that the concentration of ionized impurities is less than $10^{10}$ cm$^{-3}$. In one sample the top (001) surface was ion-implanted at room-temperature with $^{14}$N using a Danfysik 350 kV high current implanter. Three different energy levels and doses, 30 keV ($3\times10^{13}$ cm$^{-2}$), 60 keV ($4\times10^{13}$ cm$^{-2}$) and 90 keV ($3\times10^{13}$ cm$^{-2}$) were used for the implantation. This was followed by anneal at 800 °C for 2 h to create the desired NV centers. On all samples a 30 nm thick $Al_2O_3$ oxide layer was deposited on the top (001) oxygen terminated surface of the diamond using a Picosun R200 system for atomic layer deposition (ALD) from trimethylaluminum and ozone as precursors. The deposition temperature was 300 °C. The ALD process is initiated with ozone present as a last cleaning step and also to ensure high oxygen termination coverage. Openings in the oxide for the source and drain contacts were made by lithographic patterning and hydrofluoric (HF) etching. The samples were metallized by Ti/Al (20 nm/300 nm) evaporation and source, gate and drain contacts were formed by standard lithographic techniques and wet etching. The back (001) surface was metallized by evaporating 10 nm Au covering the entire surface providing an optically semitransparent back contact.

## *Experimental details:*

The samples are mounted in a customized Janis ST-300MS vacuum cryostat with UV optical access. The sample temperature was monitored using a LakeShore 331 temperature controller with a calibrated TG-120-CU-HT-1.4H

GaAlAs diode sensor in good thermal contact with the sample to keep the temperature constant within 0.1 K. The current is measured using broadband low-noise amplifiers with $R = 50$ Ω input impedance, together with a digital sampling oscilloscope (3 GHz, 10 GS/s). The trigger jitter is < 100 ps, which enables averaging over many (typically 100) pulses to improve the signal-to-noise ratio. Samples were illuminated with 800 ps (FWHM) 213 nm pulses from a CryLaS FQSS213-Q4-STA passively Q-switched DPSS laser. A 213 nm interference filter ensures that no other wavelengths are transmitted. Reflective UV optics are used for focusing, in conjunction with an EHD UK1158 UV-enhanced CCD camera for imaging and positioning. The luminescence was monitored with a Dhyana 400D Peltier-cooled sCMOS camera in combination with standard microscope optics and a longpass filter with 550 nm cut-off. The luminescence spectra were obtained using an Ocean Optics S2000 fibre optic spectrometer.

**REFERENCES**
(1) Jacoboni, C.; Reggiani, L. The Monte Carlo Method for the Solution of Charge Transport in Semiconductors with Applications to Covalent Materials. *Rev. Mod. Phys.* **1983**, *55* (3), 645.
(2) Hammersberg, J.; Majdi, S.; Kovi, K. K.; Suntornwipat, N.; Gabrysch, M.; Isberg, J. Stability of Polarized States for Diamond Valleytronics. *Appl. Phys. Lett.* **2014**, *104*, 232105.
(3) Isberg, J.; Gabrysch, M.; Tajani, A.; Twitchen, D. J. Transient Current Electric Field Profiling of Single Crystal CVD Diamond. *Semicond. Sci. Technol.* **2006**, *21* (8), 1193.